\documentclass[referee, sn-basic]{sn-jnl}% Math and Physical Sciences Reference Style

\usepackage{graphicx}%
\usepackage{multirow}%
\usepackage{amsmath,amssymb,amsfonts}%
\usepackage{amsthm}%
\usepackage{mathrsfs}%
\usepackage[title]{appendix}%
\usepackage{xcolor}%
\usepackage{textcomp}%
\usepackage{manyfoot}%
\usepackage{booktabs}%
\usepackage{algorithm}%
\usepackage{algorithmicx}%
\usepackage{algpseudocode}%
\usepackage{listings}%
\usepackage{natbib} %
\usepackage{subcaption}
\theoremstyle{thmstyleone}%
\theoremstyle{thmstyletwo}%

\theoremstyle{thmstylethree}%

\begin{document}

\title[Article Title]{Spatio-temporal evolution of surface temperature trends in Ghana (1983--2021): a multi-station approach}

%\title[Article Title]{Validation of Satellite and Reanalysis Rainfall Products in Ghana and Zambia}


%\author{\fnm{Anonymous} \sur{Author}}

\author*[1,2]{\fnm{John} \sur{Bagiliko}}\email{john.bagiliko@aims-senegal.org}

\author[3]{\fnm{David} \sur{Stern}}\email{d.a.stern@idems.international}

\author[1]{\fnm{Denis} \sur{Ndanguza}}\email{dndanguzarusatsi@ur.ac.rw}

\affil*[1]{\orgdiv{Department of Mathematics}, \orgname{School of Science, College of Science and Technology, University of Rwanda}, \orgaddress{\city{Kigali}, \postcode{P.O. Box 3900}, \country{Rwanda}}}

\affil[2]{ \orgname{African Institute for Mathematical Sciences, Research and Innovation Center}, \orgaddress{\street{Rue KG590 ST}, \city{Kigali}, \country{Rwanda}}}

\affil[3]{ \orgname{IDEMS International}, \orgaddress{\street{RG2 7AX}, \city{Reading}, \country{United Kingdom}}}


\abstract{
	Surface temperature is a fundamental Essential Climate Variable, serving as a primary indicator of climate change and exerting a profound influence on ecosystems, agriculture, and human livelihoods. Although existing research provides a foundation for understanding the climate of Ghana, there remains an opportunity to enhance this landscape with granular station-level analysis. Such high-resolution analysis complements existing studies by capturing localised climatic nuances. This study conducts a detailed spatio-temporal analysis of temperature trends across 22 meteorological stations from 1983 to 2021. Using daily maximum (Tmax) and minimum (Tmin) observations, data were subjected to quality control, homogeneity testing, and homogenisation according to World Meteorological Organisation (WMO) standards, using AgERA5 reanalysis as a reference.The significance and magnitude of trends were determined using the Modified Mann-Kendall test, which is robust in handling potential effects of autocorrelation, and Sen's slope estimator. Results revealed that temperature trends in Ghana are highly localised and seasonal, highlighting the necessity for more studies of this nature. A critical finding is the asymmetric warming across the country, with minimum temperatures rising at an accelerated rate compared to maximum temperatures. This narrowing of the diurnal temperature range poses significant threats to agricultural stability and public health because nocturnal cooling is diminished. These findings underscore the urgent need for site-specific, seasonal climate monitoring to inform customised adaptation strategies. To mitigate these impacts, the study recommends a robust policy framework focusing on afforestation and the transition to green energy.
}




\maketitle

\section{Introduction}\label{sec1}
Africa is uniquely vulnerable to the devastating impacts of climate change and has great diversity \citep{Adeyeri2026}. In Ghana, as in many other African nations, there is an urgent need for robust climate research to provide the empirical foundation necessary for informed policy-making and the development of effective mitigation strategies \citep{Adeyeri2026, Pitman2011}. 

Surface temperature is a fundamental ECV, serving as a primary indicator of climate change and exerting a profound influence on ecosystems, agriculture, and human livelihoods. Existing literature provides a vital foundation for understanding Ghana's climate (e.g \cite{Ankrah2023, Atiah2021, Arfasa2024, Asamoah2020, BaffourAta2021, Cudjoe2021, Frimpong2022, BaffourAta2023, OhenebaDornyo2022, AsareNuamah2019}). A great number of literature has predominantly focused on specific sectoral applications or restricted geographical areas (e.g \cite{Ankrah2023, Atiah2021, Arfasa2024, Asamoah2020, BaffourAta2021, Cudjoe2021, Frimpong2022, BaffourAta2023, OhenebaDornyo2022}). While nationwide assessments have been conducted, they frequently rely on spatial averaging within agroecological or climatic zones \citep{AsareNuamah2019}. However, the agroecological and climatic zoning of Ghana is a subject of ongoing debate, and some studies propose different classifications due to an evolving climate, such as \cite{Yamba2023} and \cite{Bessah2022}). The inclusion of stations within incorrectly defined zones can lead to unintended biases and masking critical localised trends. And this can in turn lead to misalignment of adaptation measures with localised risks at specific locations \citep{Adeyeri2026}. This provides an opportunity to enhance this landscape with more granular station-level analysis  in Ghana.

The current study contributes to the literature by performing a spatio-temporal analysis of temperature trends across a network of 22 stations in Ghana for the period 1983--2021. The primary objective was to evaluate whether temperature trends in Ghana are localised or if they align with existing regional zoning and spatial averaging methodologies. The second objective was to analyse the evolution of these temperature trends at different times of the year. 

The data comprised daily observations of maximum temperature (Tmax) and minimum temperature (Tmin). To ensure data integrity, all records underwent a preliminary quality control in accordance with the recommendations of the World Meteorological Organisation (WMO) \citep{WMO2021}. The homogeneity of the data was subsequently evaluated using four distinct tests, with the AgERA5 reanalysis dataset \citep{agera5} serving as a reference for homogenisation. Finally, the Modified Mann-Kendall (MMK) test and Sen's slope estimator were used to determine the significance and magnitude of the observed trends.   

Section \ref{sec_methods} describes the materials and methods used; the results are presented in Section \ref{sec2}, and discussed in Section \ref{sec:discussion}; and the conclusion is given in Section \ref{sec:conclusion}.

\section{Materials and methods}\label{sec_methods}
This section describes the geographical and climatic characteristics of the study area, the datasets used, and the methods employed in this research. They are presented in the following four subsections.
%This section provides a description of the study area, the data, and the methods used in the data. They are presented in the next four subsections.

\subsection{Study Area}
\begin{figure*}[ht] 
	\centering    \includegraphics[width=0.95\textwidth]{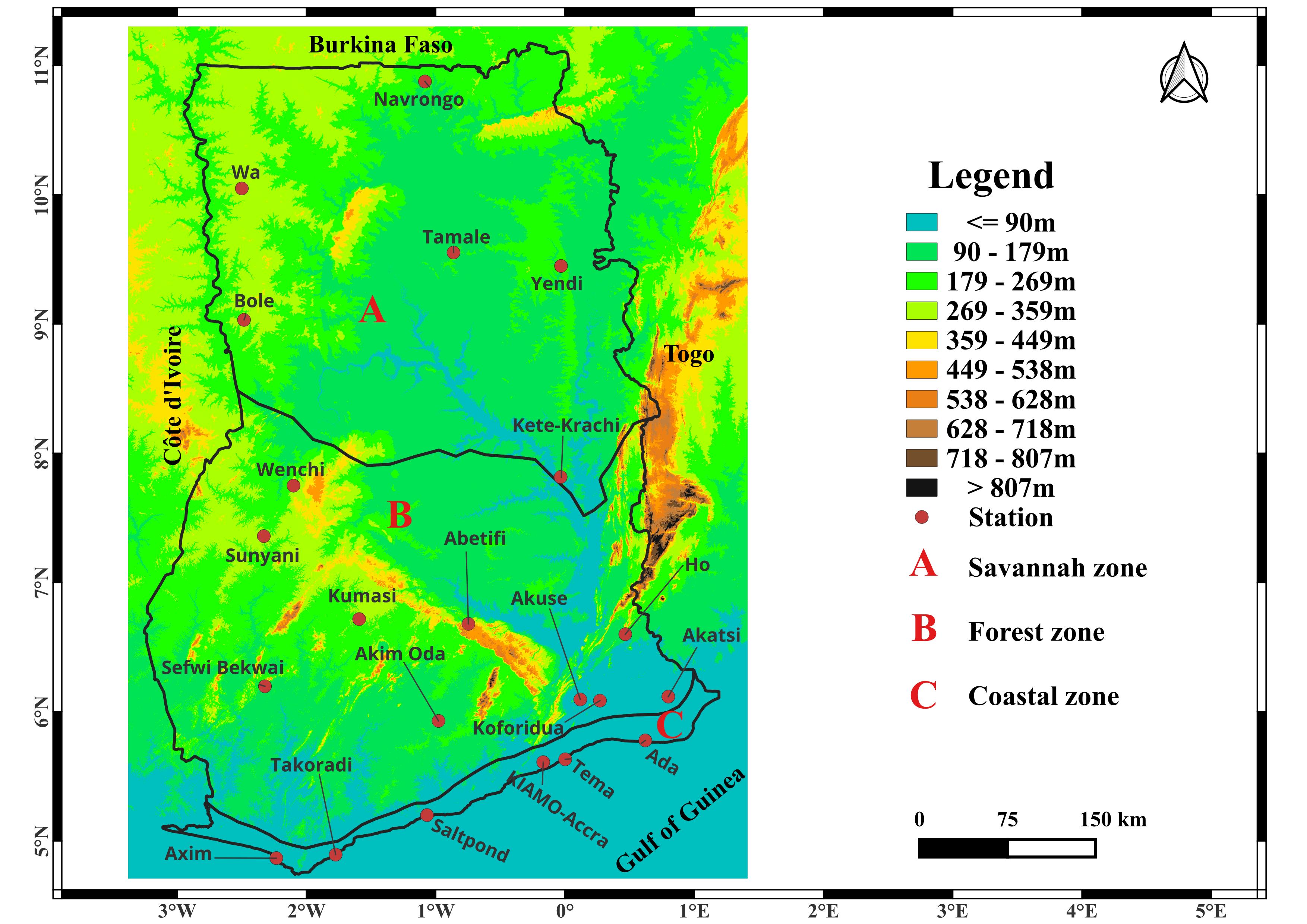}
	\caption{Physical and climatic characteristics of the study area in Ghana. Climatic zoning is defined according to \citet{Bessah2022}, with topographical elevations derived from the Shuttle Radar Topography Mission (SRTM) Digital Elevation Model \citep{Farr2007}}
	\label{fig:study_area_map}
\end{figure*}

\begin{figure*}[ht] 
	\centering    \includegraphics[width=0.95\textwidth]{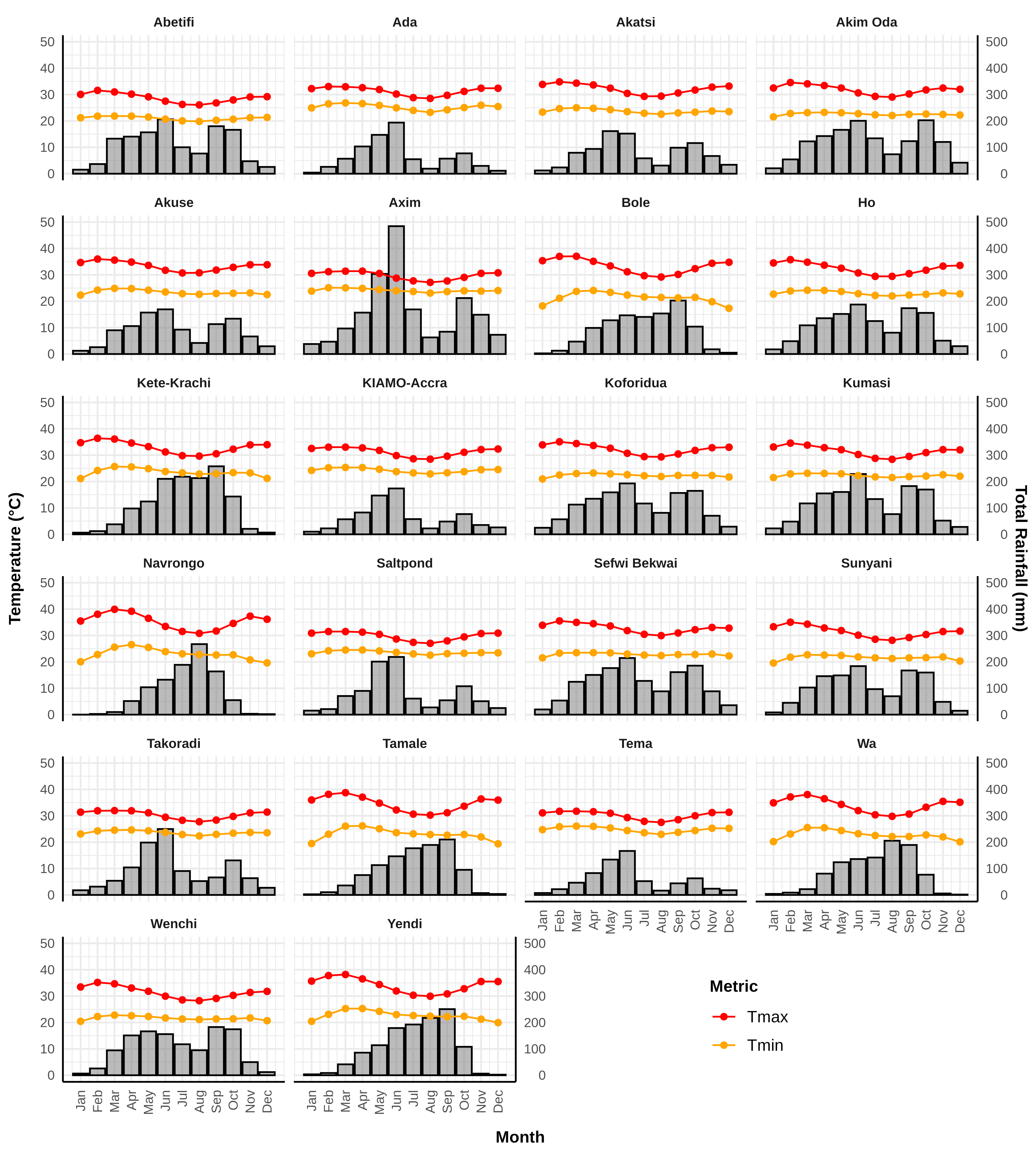}
	\caption{Ombrothermic diagrams of the stations considered in the study. The bars show the mean monthly total rainfall values while the line plots show the mean Tmax (red) and mean Tmin (orange). The rainfall scale $=$ temperature scale $\times$ 10}
	\label{fig:ombrothermic_diag}
\end{figure*}

This study is based on 22 stations in Ghana (Figure~\ref{fig:study_area_map}). These stations are located in regions with varied topographic characteristics, ranging from high-altitude locations such as Abetifi to low-lying coastal areas such as Ada. Stations located in the Savannah zone have uni-modal rainfall patterns, usually starting in April, peaking around August and September, and ending around October (Figure~\ref{fig:ombrothermic_diag}). Their driest months are usually January and December. The average minimum temperatures are generally above 20 $^\circ$C, with minimum values usually in January and December (the driest months), peaking around March, and lowering gently throughout the rainy season (Figure~\ref{fig:ombrothermic_diag}). The average maximum temperatures are generally above 30 $^\circ$C and peak around March at about 40 $^\circ$C, trough around July, August, and peak again in November around 35 $^\circ$C (Figure~\ref{fig:ombrothermic_diag}). The stations in the Forest zone as well as the Coastal zone have bimodal rainfall patterns, with peaks around June and October, respectively. Their driest months are also usually January and December. The average minimum temperatures are relatively constant throughout the year, around or above 20 $^\circ$C. Although maximum temperatures are relatively constant, compared to those in the Savannah zone, they peak around February and around November, and trough around the dry period between the rainy seasons, August (Figure~\ref{fig:ombrothermic_diag}). The diurnal range (Tmax - Tmin) is generally highest in January and December for stations in all climatic zones and lowest in August. It is highest and more seasonal for the stations in the Savannah zone, followed by the stations in the Forest zone. It is lowest and relatively constant throughout the year for stations in the Coastal zone.

\subsection{Data and quality control}
The temperature data was subjected to quality control procedures in accordance with the WMO recommendations. Several internal consistency verifications were performed to ensure the integrity of the data. First, we verified that the minimum temperature (Tmin) did not exceed the maximum temperature (Tmax) for any given day. In instances where this occurred, the values were frequently found to be swapped; such entries were swapped accordingly. In other cases, obvious typographical errors, such as 32.1 written as 23.1, were corrected by cross-referencing adjacent data points. If the discrepancy could not be resolved by these methods, the values were flagged as missing. Threshold verifications were also applied to identify unrealistic extremes. Maximum temperature values exceeding 50 $^\circ$C were investigated, often revealing digit reversals such as 63 instead of 36. Similarly, minimum temperatures below 10 $^\circ$C were examined, revealing decimal placement errors such as 2.4 instead of 24. These were corrected on the basis of the context of neighbouring values. To address temporal consistency, we monitored large fluctuations between consecutive days and the occurrence of persistent identical values. Inter-diurnal changes in Tmax or Tmin exceeding 10 $^\circ$C were scrutinised for typos or parameter swaps. Furthermore, if a constant value was recorded for more than five consecutive days, those observations were deemed suspicious and marked as missing to avoid biassing the trend analysis.

\subsection{Homogeneity test and homogenisation}
The next phase of the analysis involved conducting homogeneity tests on the data set, a critical step recommended by the World Meteorological Organisation (WMO). This procedure ensures that any detected trends are representative of genuine climatic shifts rather than non-climatic artefacts, such as changes in station location, instrument calibration, or land-use alterations. Homogeneity testing was specifically applied to the monthly and annual mean Diurnal Temperature Range (DTR) series, as these are notably more sensitive to inconsistencies that may remain undetected in the individual Tmax or Tmin series \citep{wijngaard2003homogeneity}.

Four statistical tests were used: the Standard Normal Homogeneity Test (SNHT) \citep{alexandersson1986homogeneity}, Pettitt's Test (PT) \citep{pettitt1979non}, Buishand's Likelihood Ratio Test (BLRT) \citep{Buishand1984}, and Buishand's U Test (BUT) \citep{Buishand1984}. These tests possess complementary strengths (further detailed in Appendix~\ref{secA1}). The null hypothesis ($H_{0}$) for each test assumed a homogeneous series, while the alternative hypothesis ($H_{a}$) assumed an inhomogeneous series. Statistical significance was set at a confidence level of 95\%; thus, a p-value less than 0.05 resulted in the rejection of $H_{0}$. These tests were performed using the Python library \textit{pyHomogeneity} \citep{pyHomogeneity}, with 20,000 Monte Carlo simulations used for each test to estimate the p-value.

Following the methodology of \cite{wijngaard2003homogeneity}, stations were categorised based on the test results: class A ("Useful"), zero or one test rejected $H_{0}$; class B ("Doubtful"), two tests rejected $H_{0}$; and class C ("Suspect"), three or all four tests rejected $H_{0}$. This classification framework provided an objective measure of the reliability of subsequent trend analyses. %These results were mapped spatially, allowing for an evaluation of whether observed inhomogeneities were isolated or related to broader regional patterns.

The Tmax and Tmin series were homogenised using the \textit{Climatol} package \cite{climatol} in R. This package was selected for its robust handling of missing values and comprehensive documentation. Daily estimates from the nearest AgERA5 pixel to each station were used to serve as reference series. The homogenisation process was first performed on monthly aggregates, which were subsequently used to adjust the daily records. AgERA5 was chosen due to its high spatial resolution (0.1$^\circ$) and accessibility. While reanalysis data may contain systematic biases, they are generally considered more spatially and temporally homogeneous than raw station records, making them suitable references for this procedure \citep{climatol}.

\subsection{Autocorrelation adjustment and trend estimation}
To analyse the monthly and annual temperature trends across the study area, we employed the Modified Mann-Kendall (MMK) test \citep{Hamed1998} in conjunction with Sen's slope estimator \citep{Sen1968}. The original Mann-Kendall (MK) test statistic $S$ is defined as:

\begin{equation}
	S = \sum_{i=1}^{n-1} \sum_{j=i+1}^{n} \text{sgn}(x_j - x_i)
\end{equation}

where $n$ is the number of data points, $x_i$ and $x_j$ are the sequential data values, and the sign function $\text{sgn}(x_j - x_i)$ is given by:

\begin{equation}
	\text{sgn}(x_j - x_i) = 
	\begin{cases} 
		1 & \text{if } (x_j - x_i) > 0 \\
		0 & \text{if } (x_j - x_i) = 0 \\
		-1 & \text{if } (x_j - x_i) < 0 
	\end{cases}
\end{equation}

For datasets where $n > 10$, the variance of $S$ is calculated to determine the standardised test statistic $Z$:

\begin{equation}
	\text{Var}(S) = \frac{n(n-1)(2n+5) - \sum_{k=1}^{m} t_k(t_k-1)(2t_k+5)}{18}
\end{equation}

where $m$ is the number of linked groups and $t_k$ is the number of data points in the $k^{th}$ group. The standardised statistic $Z$ (\ref{z_def}) follows a normal distribution, with positive and negative values indicating increasing and decreasing trends, respectively. The standardised $Z$ is given by:

\begin{equation} \label{z_def}
	Z =
	\begin{cases}
		\frac{S-1}{\sqrt{\text{Var}(S)}} & \text{if } S > 0 \\
		0 & \text{if } S = 0 \\
		\frac{S+1}{\sqrt{\text{Var}(S)}} & \text{if } S < 0
	\end{cases}
\end{equation}

The decision to reject or accept the null hypothesis is then
made by comparing the Z value with the critical value at a chosen level
of significance.

While MK test is a robust non-parametric method for trend detection, it is highly sensitive to serial correlation. In climatic time series, autocorrelation can lead to an inflation of the $S$ statistic variance, resulting in the false detection of significant trends \citep{Hamed1998}.

The MMK test addresses this limitation by adjusting the variance of the MK statistic $S$. The original variance, $Var(S)$, is modified to a corrected variance, $Var^*(S)$, to account for the autocorrelation structure of the data series:

\begin{equation}
	Var^*(S) = Var(S) \cdot \frac{n}{n^*} = Var(S) \left[ 1 + \frac{2}{n(n-1)(n-2)} \sum_{i=1}^{n-1} (n-i)(n-i-1)(n-i-2)\rho_i \right]
\end{equation}

where $n$ represents the number of observations and $\rho_i$ denotes the autocorrelation function of the ranks of the data. This modification ensures that the standardised test statistic $Z$ remains reliable even when the assumption of independence is violated \citep{Hamed1998}.

The magnitude and direction of the trends were estimated using Sen's slope estimator. This method is particularly well-suited for climatological data as it is non-parametric, robust against outliers, and does not require data to follow a normal distribution, unlike ordinary least-squares (OLS) regression \citep{Sen1968}. 

Statistical computations for both the Modified Mann-Kendall (MMK) test and Sen's slope were performed using the \textit{mmkh} function within the \textit{modifiedmk} R package. All trend analyses were conducted at a 95\% confidence level; the null hypothesis ($H_0$) of no trend was rejected if the calculated $p$-value was less than 0.05.

\section{Results}\label{sec2}
This section presents the findings of the study. The results of the homogeneity assessments are detailed in Subsection~\ref{res_homogen}, followed by a comprehensive evaluation of the spatio-temporal temperature trends in Subsection~\ref{res_trends}.
\subsection{Homogeneity Test Results}\label{res_homogen}
% Table \ref{tab:homogeneity_results} presents the results of the homogeneity tests performed in the annual DTR series at all stations. Of the 22 stations, 8 were found to be in class A, meaning that their series were found to be homogeneous, 1 was found to be in class B, while the remaining 13 were classified in class C (meaning that these stations were found to be "suspect stations"). Apart from Akatsi and Kete-Krachi, the rest of the class A stations had all tests unanimously failing to reject $H_{o}$. And also, apart from Ho and Ada, the other stations in class C had all tests unanimously rejecting $H_{o}$. However, further analysis showed that the change points were usually not the same years between tests (see Appendix \ref{secA2}), mainly relating the unique properties of these tests (see Appendix \ref{secA1}).   

The results of the homogeneity tests for the annual DTR series are summarised in Table~\ref{tab:homogeneity_results}. Of the 22 stations analysed, eight were classified as Class A (Useful), one as Class B (Doubtful), and 13 as Class C (Suspect). Within Class A, the majority of stations unanimously failed to reject $H_{0}$, with the exception of Akatsi and Kete-Krachi. Conversely, nearly all Class C stations exhibited a unanimous rejection of $H_{0}$ across all four tests, excluding Ho and Ada. In particular, the detected change points frequently varied between tests (Figures~\ref{fig:snht_results}--\ref{fig:but_results}), attributable to the specific statistical properties and sensitivities of the individual tests (Appendix~\ref{secA1}).

\begin{figure}[H] % [p] places it on its own page if it's large
	\centering
	\includegraphics[width=\textwidth]{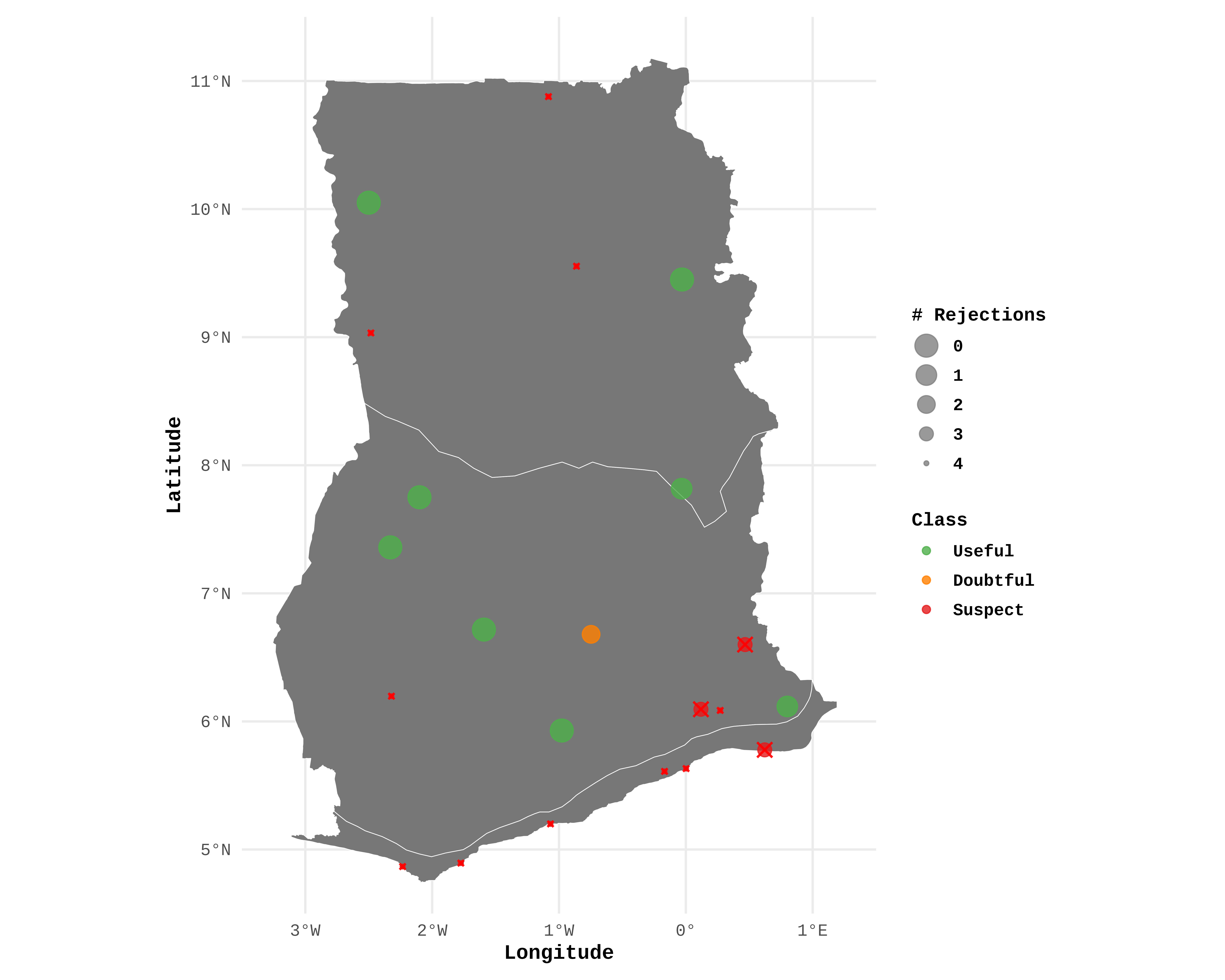}
	\caption{A map showing the homogeneity test results for the annual DTR series across all stations. Green points represent "Useful" stations, oranges "Suspect" stations, while red points with an X on them represent "Suspect" stations. The size of the points depicts, in the reverse order, the number of tests that rejected $H_{o}$}
	\label{fig:annual_homogen}
\end{figure}

The spatial mapping of these results (Figure~\ref{fig:annual_homogen}) revealed that while the Class A stations are distributed throughout the Savannah and Forest zones, a distinct geographical pattern emerges for Class C. In particular, every station located in the southernmost part of the country, particularly within the Coastal zone, was classified as "Suspect."

\begin{sidewaystable}
	\centering
	\caption{Homogeneity test results for the annual DTR series across all stations. Values represent test statistics and associated p-values}
	\label{tab:homogeneity_results}
	\small 
	\begin{tabular*}{\textheight}{@{\extracolsep{\fill}}lcccccccccc}
		\toprule
		Station & $U$ (PT) & $p$ (PT) & $T$ (SNHT) & $p$ (SNHT) & $V$ (BLRT) & $p$ (BLRT) & $U$ (BUT) & $p$ (BUT) & Rejects & Class \\
		\midrule
		Abetifi      & 87  & 0.263 & 13.870 & 0.0004 & 0.692 & 0.0007 & 0.255 & 0.182 & 2 & B (Doubtful) \\
		Ada          & 103 & 0.024 & 8.091  & 0.0345 & 0.569 & 0.0351 & 0.435 & 0.052 & 3 & C (Suspect) \\
		Akatsi       & 124 & 0.051 & 6.745  & 0.0930 & 0.474 & 0.0954 & 0.457 & 0.046 & 1 & A (Useful) \\
		Akim Oda     & 117 & 0.054 & 6.663  & 0.0962 & 0.479 & 0.0950 & 0.341 & 0.102 & 0 & A (Useful) \\
		Akuse        & 88  & 0.162 & 17.150 & 5.0e-5 & 0.796 & 1.0e-4 & 0.468 & 0.045 & 3 & C (Suspect) \\
		Axim         & 140 & 0.002 & 11.483 & 0.0027 & 0.652 & 0.0032 & 0.951 & 0.002 & 4 & C (Suspect) \\
		Bole         & 132 & 0.010 & 15.243 & 1.0e-4 & 0.738 & 1.0e-4 & 0.734 & 0.007 & 4 & C (Suspect) \\
		Ho           & 102 & 0.127 & 11.016 & 0.0044 & 0.616 & 0.0049 & 0.473 & 0.039 & 3 & C (Suspect) \\
		KIAMO-Accra  & 185 & 0.000 & 16.822 & 0.0000 & 0.762 & 5.0e-5 & 1.810 & 0.000 & 4 & C (Suspect) \\
		Kete-Krachi  & 122 & 0.041 & 6.187  & 0.1291 & 0.462 & 0.1227 & 0.279 & 0.152 & 1 & A (Useful) \\
		Koforidua    & 154 & 0.012 & 8.062  & 0.0445 & 0.510 & 0.0436 & 0.639 & 0.014 & 4 & C (Suspect) \\
		Kumasi       & 97  & 0.352 & 3.016  & 0.6173 & 0.302 & 0.6053 & 0.148 & 0.405 & 0 & A (Useful) \\
		Navrongo     & 146 & 0.030 & 10.642 & 0.0088 & 0.577 & 0.0093 & 0.705 & 0.010 & 4 & C (Suspect) \\
		Saltpond     & 140 & 0.029 & 9.448  & 0.0196 & 0.552 & 0.0188 & 0.661 & 0.012 & 4 & C (Suspect) \\
		Sefwi Bekwai & 146 & 0.011 & 10.773 & 0.0064 & 0.599 & 0.0071 & 0.585 & 0.022 & 4 & C (Suspect) \\
		Sunyani      & 104 & 0.231 & 6.010  & 0.1470 & 0.433 & 0.1488 & 0.138 & 0.443 & 0 & A (Useful) \\
		Takoradi     & 182 & 1.0e-4& 14.600 & 2.0e-4 & 0.721 & 5.0e-5 & 1.584 & 0.000 & 4 & C (Suspect) \\
		Tamale       & 156 & 1.0e-4& 15.600 & 5.0e-5 & 0.774 & 1.0e-4 & 1.501 & 0.000 & 4 & C (Suspect) \\
		Tema         & 265 & 0.000 & 17.819 & 5.0e-5 & 0.735 & 1.0e-4 & 1.000 & 0.001 & 4 & C (Suspect) \\
		Wa           & 68  & 0.657 & 2.856  & 0.6375 & 0.304 & 0.6398 & 0.106 & 0.572 & 0 & A (Useful) \\
		Wenchi       & 52  & 0.101 & 2.734  & 0.5937 & 0.390 & 0.5976 & 0.183 & 0.314 & 0 & A (Useful) \\
		Yendi        & 74  & 0.112 & 5.343  & 0.1728 & 0.482 & 0.1752 & 0.255 & 0.181 & 0 & A (Useful) \\
		\bottomrule
	\end{tabular*}
\end{sidewaystable}

\begin{figure}[H] 
	\centering    \includegraphics[width=0.95\textwidth]{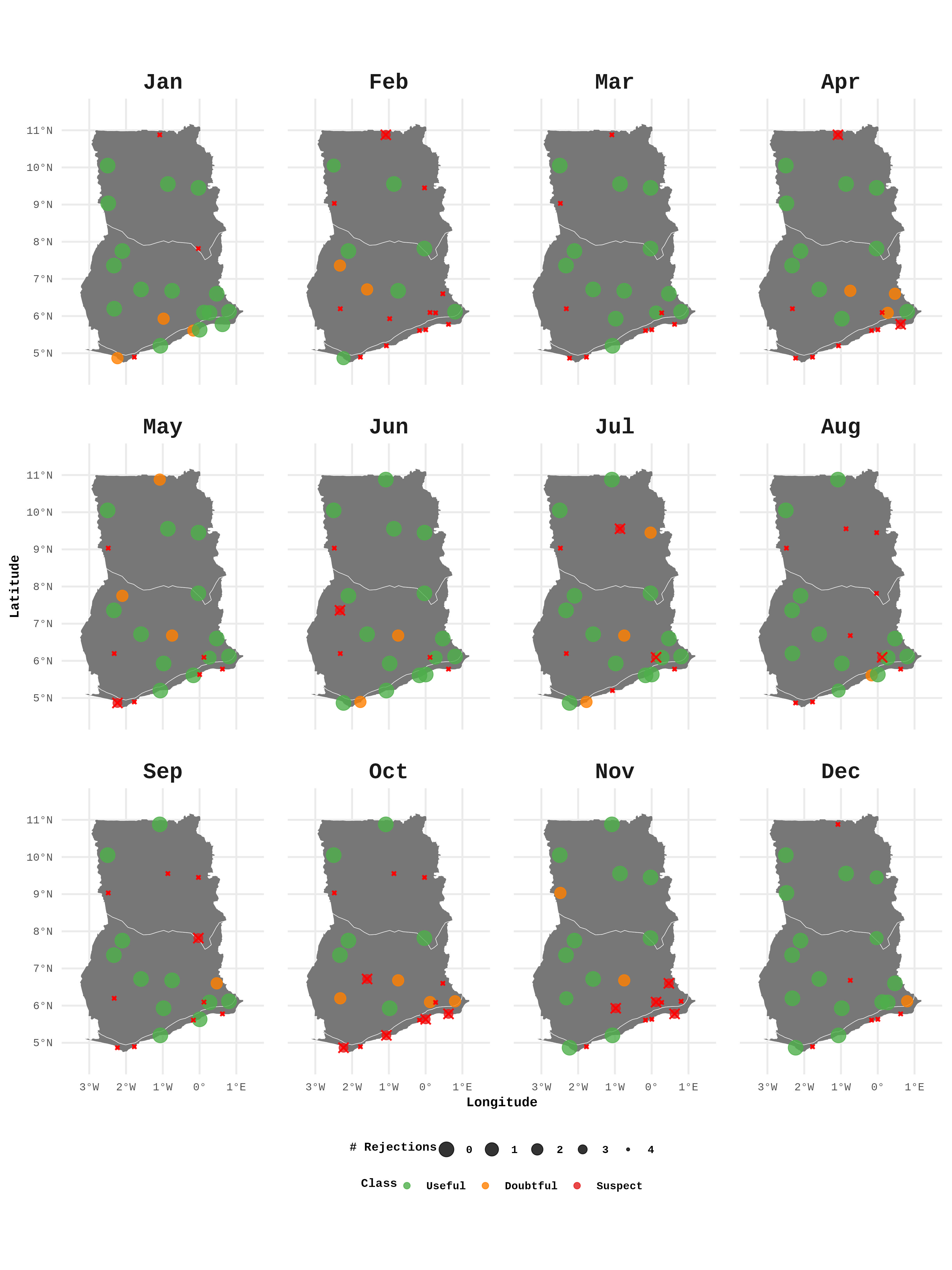}
	\caption{A map showing the homogeneity test results for the monthly DTR series across all stations. Green points represent "Useful" stations, oranges "Suspect" stations, while red points with an X on them represent "Suspect" stations. The size of the points depicts, in the reverse order, the number of tests that rejected $H_{o}$}
	\label{fig:monthly_homogen}
\end{figure}

The monthly homogeneity analysis (Figure~\ref{fig:monthly_homogen}) further supports this seasonal and spatial complexity. January and December exhibited the highest stability, with 16 and 15 stations, respectively, categorised as Class A. In contrast, February and October showed the highest levels of inhomogeneity, with 13 and 12 stations classified as Class C. A significant observation is that in the Coastal zone, all stations were classified as Class C during April and October, and all but one remained in this category during February and March.

The spatial and temporal distributions in the homogeneity results suggest that, while some inhomogeneities may be due to instrumental or procedural errors, some may actually be due to underlying climatic signals. In the absence of detailed station metadata to verify relocation or equipment changes, the interpretation of homogeneity results requires caution \citep{wijngaard2003homogeneity}. 

\subsection{Spatio-temporal analysis of temperature trends} \label{res_trends}
\begin{figure}[H]
	\centering
	\includegraphics[width=\textwidth]{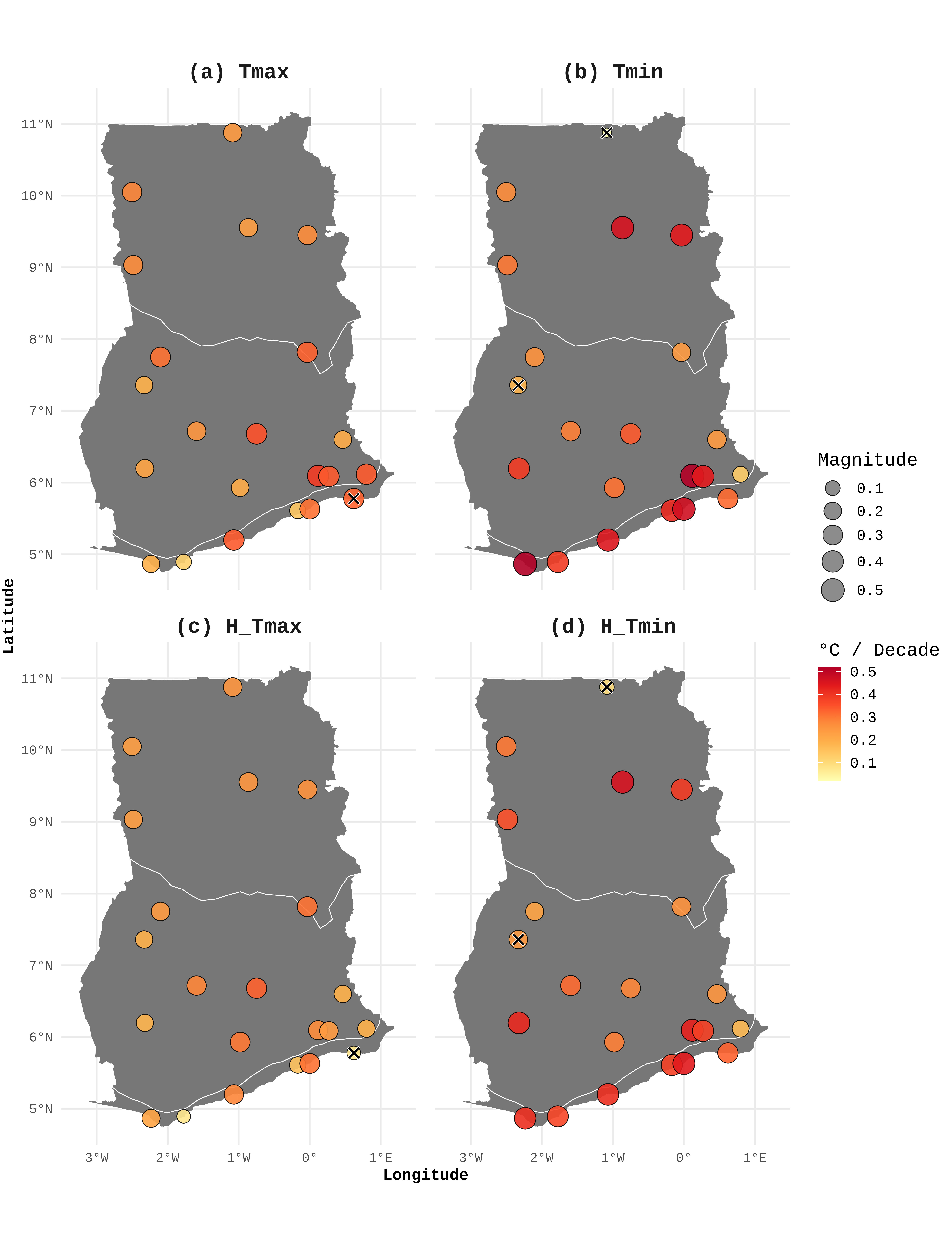}
	\caption{Spatial trends in annual maximum (Tmax) and minimum (Tmin) temperatures across Ghana. Points represent station locations, with colour indicating the trend magnitude ($^\circ$C per decade) and size representing the absolute magnitude of change. Stations where the trend is not statistically significant ($p \geq 0.05$) are overlaid with a black cross ($\times$). The y-axis represents longitudes, and the x-axis represents latitudes. The plots (a) and (b) are the original Tmax and Tmin respecctively while (c) and (d) are the homogenised Tmax and Tmin respectively}
	\label{fig:ghana_temp_trends}
\end{figure}

The results of the mean annual maximum and minimum temperature (a) and (b), respectively, as well as the homogenised Tmax (c) and Tmin (d) are shown in Figure \ref{fig:ghana_temp_trends}. Statistically significant positive trends were observed at most of the stations for both variables. Although the homogenised series showed some adjustments in the magnitude of the trends in some places, the direction and significance of the trends remained similar at all stations, meaning that the inhomogeneities did not have a high impact on the direction and significance of the trends in the annual Tmax and Tmin. The magnitude and significance of trends was observed to be localised. For example, Navrongo did not show a significant trend in Tmin, while all other stations in that zone revealed significant increase trends. For stations with significant trends, the magnitudes varied in space. E.g., Tamale revealed a warming rate of 0.5 $^\circ$C per decade while Wa showed a warming rate of approximately 0.3 $^\circ$C. Tmax revealed significant increasing trends in the range of 0.1$^\circ$C to 0.3 $^\circ$C of warming per decade (Figure~\ref{fig:ghana_temp_trends}(c)) while Tmin was in the range of 0.1 $^\circ$C to 0.5 $^\circ$C degrees of warming per decade. The results showed that Tmin had a faster rate of warming than Tmax, especially the stations in the Coastal zone and some of the Savannah and Forest locations.

\begin{figure}[H] 
	\centering    \includegraphics[width=0.95\textwidth]{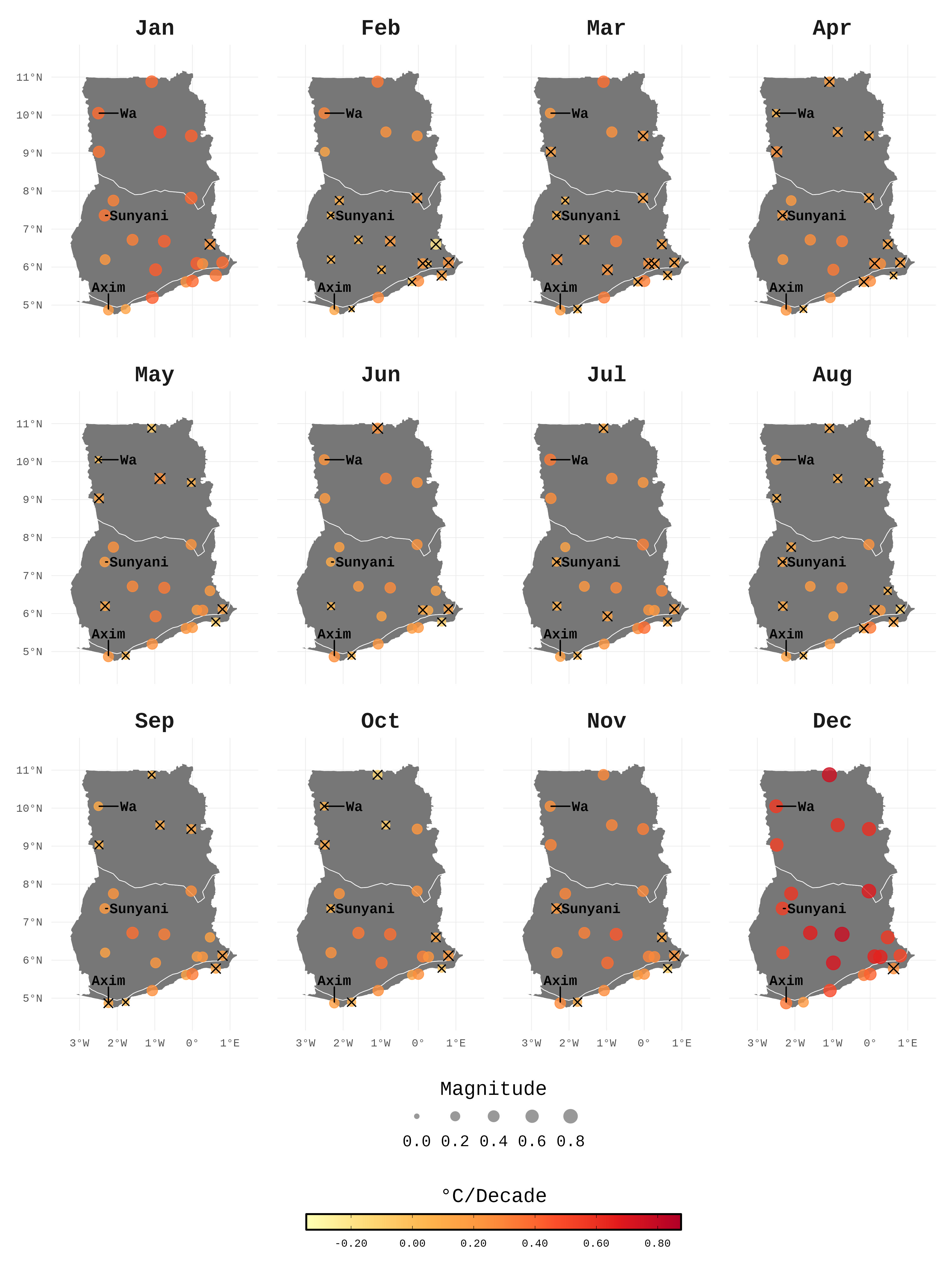}
	\caption{Monthly spatial distribution of maximum temperature (Tmax) trends across Ghana. Each panel represents a calendar month. Point colour indicates the decadal warming rate ($^\circ$C/decade), while point size reflects the absolute magnitude of change. Stations where the trend is not statistically significant ($p \geq 0.05$) are overlaid with a black cross ($\times$). The y-axis represents longitudes, and the x-axis represents latitudes}
	\label{fig:monthly_tmax_trends}
\end{figure}

\begin{figure}[H] 
	\centering
	\includegraphics[width=0.95\textwidth]{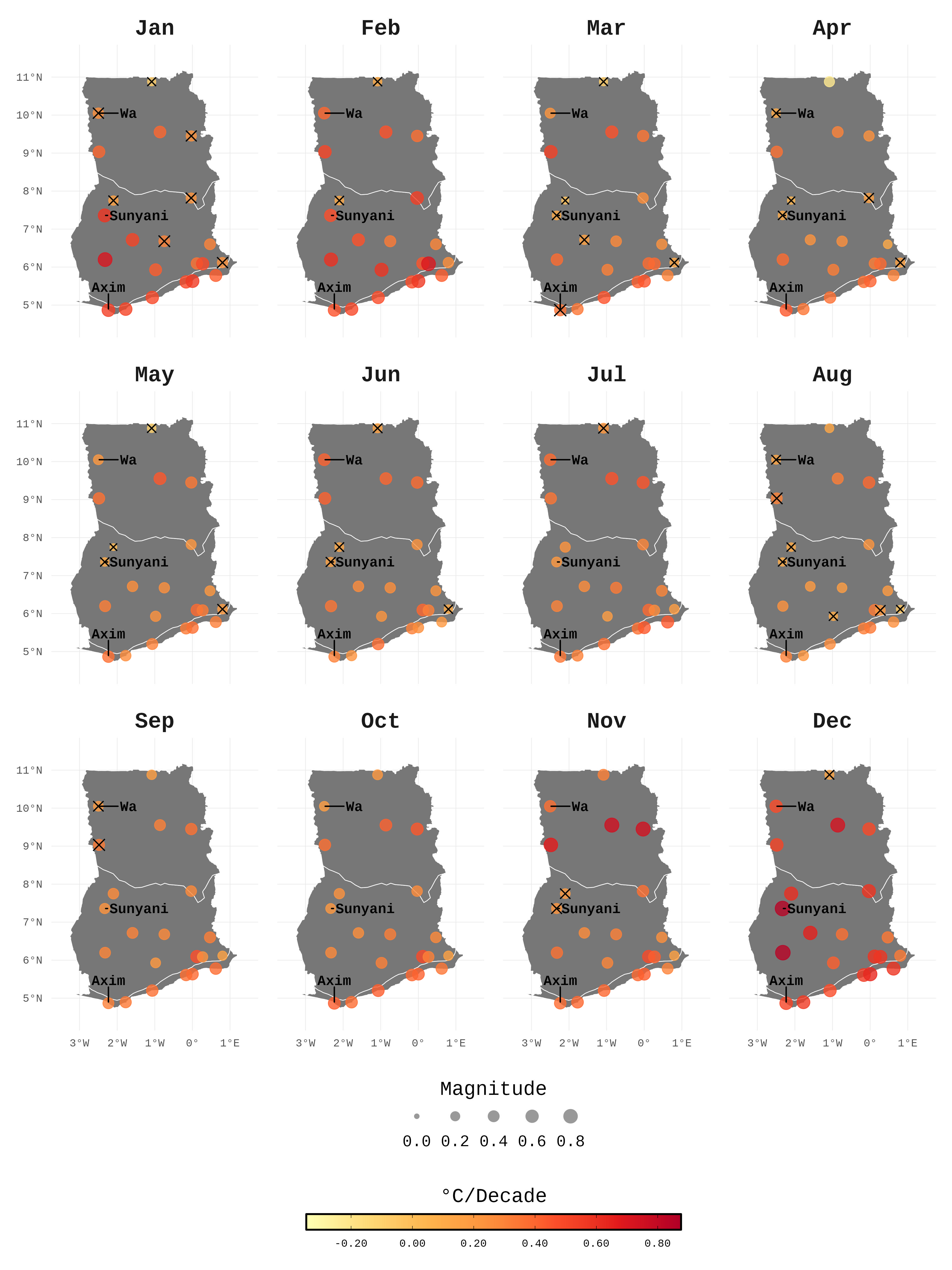}
	\caption{Monthly spatial distribution of Tmin trends across Ghana. Each panel represents a calendar month. Point colour indicates the decadal warming rate ($^\circ$C/decade), while point size reflects the absolute magnitude of change. Stations where the trend is not statistically significant ($p \geq 0.05$) are overlaid with a black cross ($\times$). The y-axis represents longitudes, and the x-axis represents latitudes}
	\label{fig:monthly_tmin_trends}
\end{figure}

Figures~\ref{fig:monthly_tmax_trends} and \ref{fig:monthly_tmin_trends} represent the long-term mean monthly trends for Tmax and Tmin, respectively (1983 - 2021) of the homogenised series. The results revealed varying trends in space and time. In some months, almost similar observations were made for most stations in the same zone, while in other months the trends were more localised. 

For Tmax, December was found to have the fastest rate of warming in maximum temperatures for most stations (between 0.2 $^\circ$C and 0.6 $^\circ$C per decade) followed by January (between 0.2 $^\circ$C and 0.4 $^\circ$C per decade). In February, when the average maximum temperatures are highest at the Forest stations (Figure~\ref{fig:ombrothermic_diag}), no significant trends were observed for these locations (Figure~\ref{fig:monthly_tmax_trends}). In April, when the rainy season usually starts in the Savannah stations, no significant trends were observed for these locations (Figure~\ref{fig:monthly_tmax_trends}). In some months, the trends and significance are much more localised, where some stations revealed significant upward trends in Tmax while no significant trends were observed for other stations, even within the same zone.

\begin{figure}[H] 
	\centering
	\includegraphics[width=0.95\textwidth]{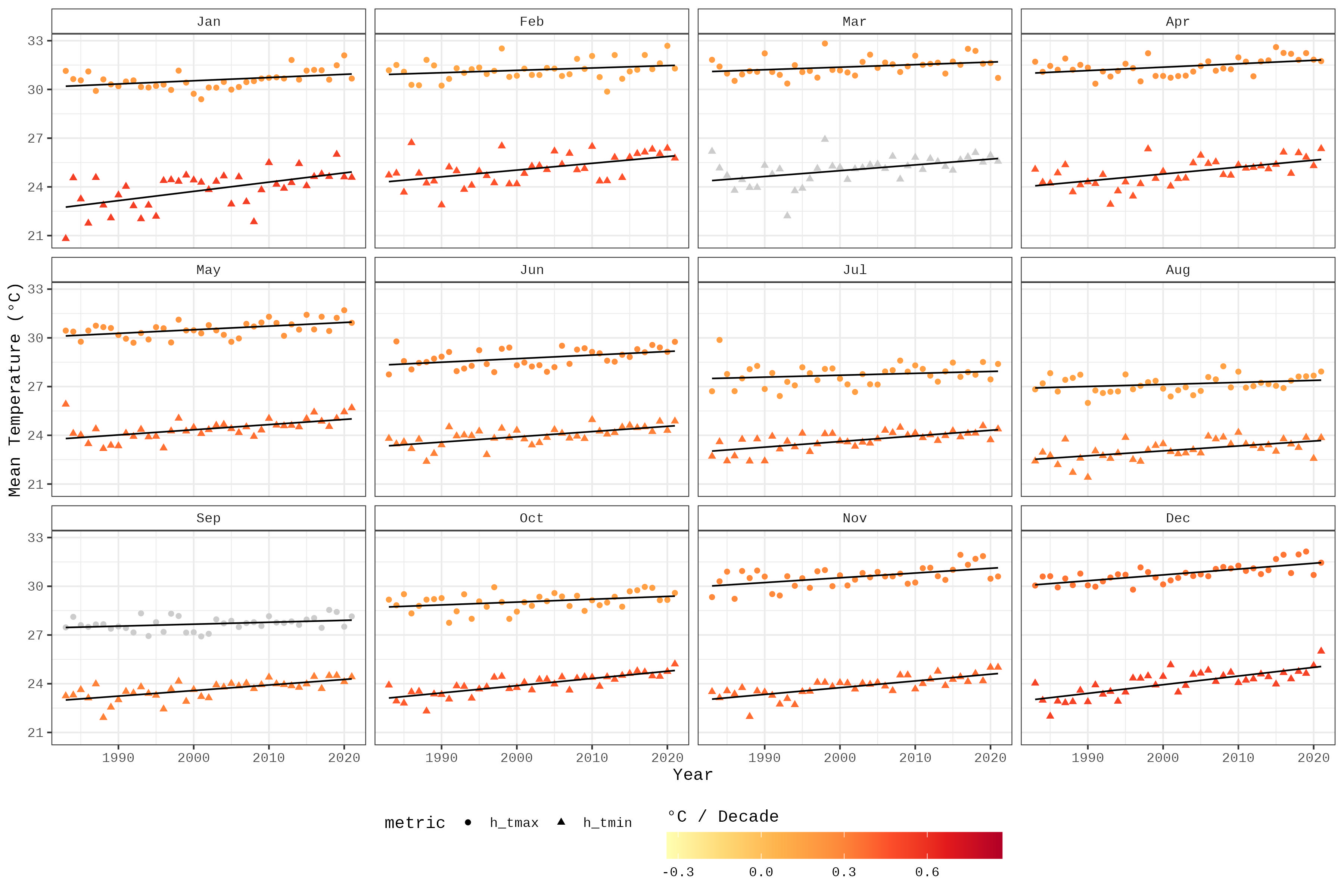}
	\caption{Monthly temporal distribution of Tmin and Tmax trends at Axim (coastal station).The y-axes represent mean Tmax (circles) and Tmin (triangles) while the x-axes represent the individual years. Grey symbols denote trends are not significant at $p < 0.05$. Black bar is the rate of change in degrees per decade}
	\label{fig:monthly_trends_axim}
\end{figure}

\begin{figure}[H] 
	\centering
	\includegraphics[width=0.95\textwidth]{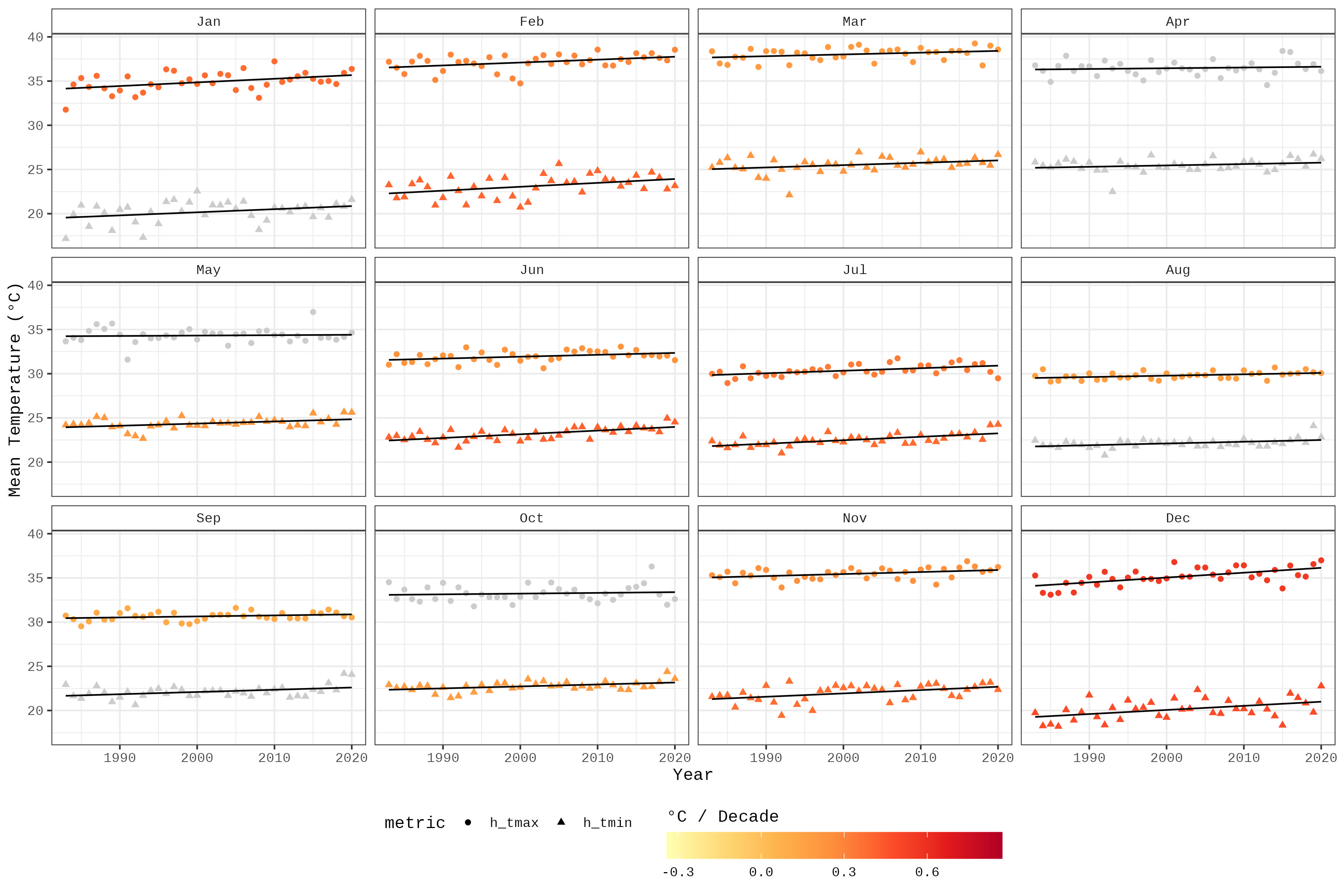}
	\caption{Monthly temporal distribution of Tmin and Tmax trends at Wa.The y-axes represent mean Tmax (circles) and Tmin (triangles) while the x-axes represent the individual years. Grey symbols denote trends are not significant at $p < 0.05$. Black bar is the rate of change in degrees per decade}
	\label{fig:monthly_trends_wa}
\end{figure}

For Tmin, the behaviour of trends was more localised. December, January and February revealed a faster rate of increase for most of the stations (between 0.4 $^\circ$C and 0.8 $^\circ$C per decade), especially in the Forest and Coastal zones (Figure~\ref{fig:monthly_tmin_trends}). The magnitude and significance of trends were seen to vary for different stations in different months. For example, Axim maintained a consistent significant increase in the trends of Tmin (between 0.3 $^\circ$C and 0.4 $^\circ$C per decade) for all months, except in March (Figures~\ref{fig:monthly_tmin_trends} and \ref{fig:monthly_trends_axim}). Sunyani had low or non significant trends for most of the months but showed a steep upward trend close to 1 $^\circ$C per decade in December followed by January and March. However, Wa revealed low but significant trends in most months, with a less steep upward trend in December (Figures~\ref{fig:monthly_tmin_trends} and \ref{fig:monthly_trends_wa}). 

Tmin was seen to increase faster in most months for most stations than Tmax, which had low or no significant trends for many of the months. In general, these results highlight the localised nature of temperature trends in Ghana. In addition, they also highlight the asymmetric nature of warming.  

\begin{figure}[H] 
	\centering
	\includegraphics[width=0.95\textwidth]{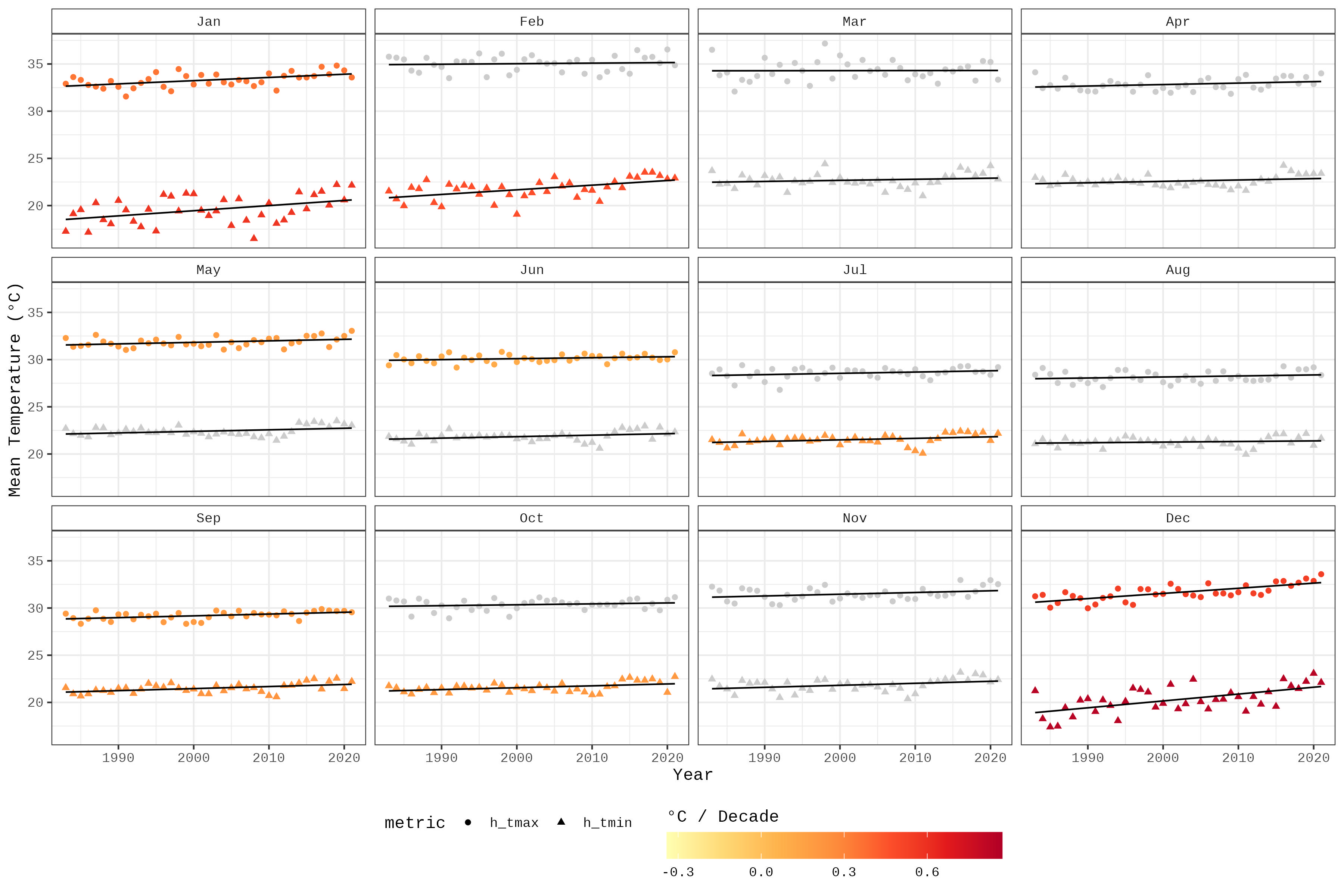}
	\caption{Monthly temporal distribution of Tmin and Tmax trends at Sunyani.The y-axes represent mean Tmax (circles) and Tmin (triangles) while the x-axes represent the individual years. Grey symbols denote trends are not significant at $p < 0.05$. Black bar is the rate of change in degrees per decade}
	\label{fig:monthly_trends_sunyani}
\end{figure}

\begin{figure}[H] % [p] places it on its own page if it's large
	\centering
	\includegraphics[width=\textwidth]{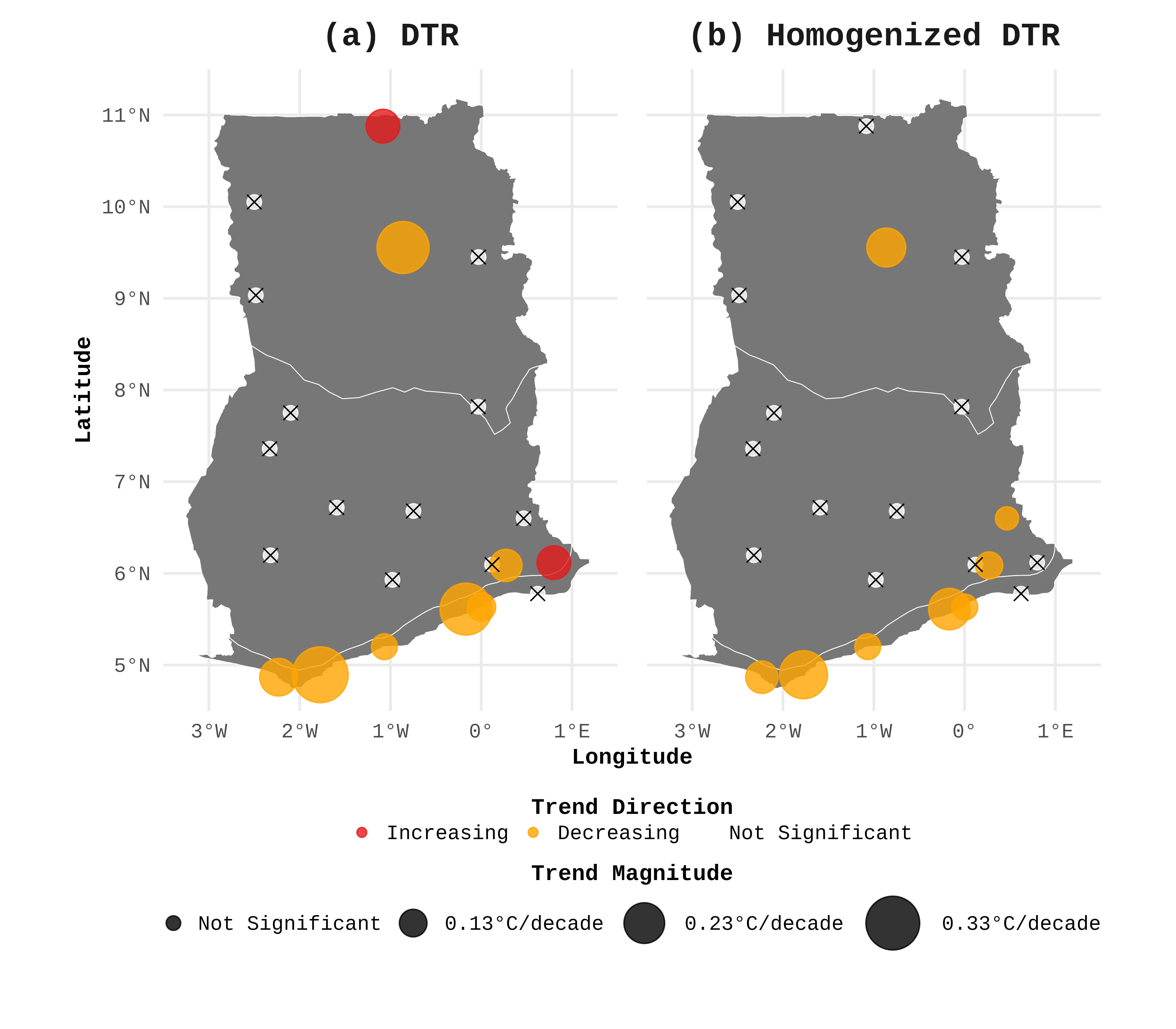}
	\caption{Mean annual diurnal temperature range (DTR)  trends of the original series (a) and homogenised series (b). Increasing trends are shown with red circles, while decreasing trends are in violet. Black X on a point indicates the trend at the location is not statistically significant (p $\geq$ 0.05). Circle size is proportional to the magnitude of change}
	\label{fig:annual_dtr_trends}
\end{figure}

Figure \ref{fig:annual_dtr_trends} shows the long-term trends in the annual diurnal temperature range (DTR) for the non-homogenised series (Figure~\ref{fig:annual_dtr_trends}(a)) and the homogenised series (Figure~\ref{fig:annual_dtr_trends}(b)). The results showed that increasing diurnal trends at two locations in the non-homogenised series were found to be non-significant trends after homogenisation. However, the direction and significance of the trends in the rest of the locations remained similar, with some adjustments in the magnitude of the trends in some places. Apart from Tamale, which revealed a significant downward trend in DTR of approximately 0.20 $^\circ$C per decade, all stations in the Savannah zone showed non-significant trends while only Ho and Koforidua were the stations in the Forest zone with significant trends with approximately 0.1 $^\circ$C per decade. With the exception of Ada, all stations in the Coastal zone revealed significant downward trends between 0.1 and 0.2 $^\circ$C per decade, but the magnitude of the trends varied per station.

\begin{figure}[H] 
	\centering
	\includegraphics[width=0.95\textwidth]{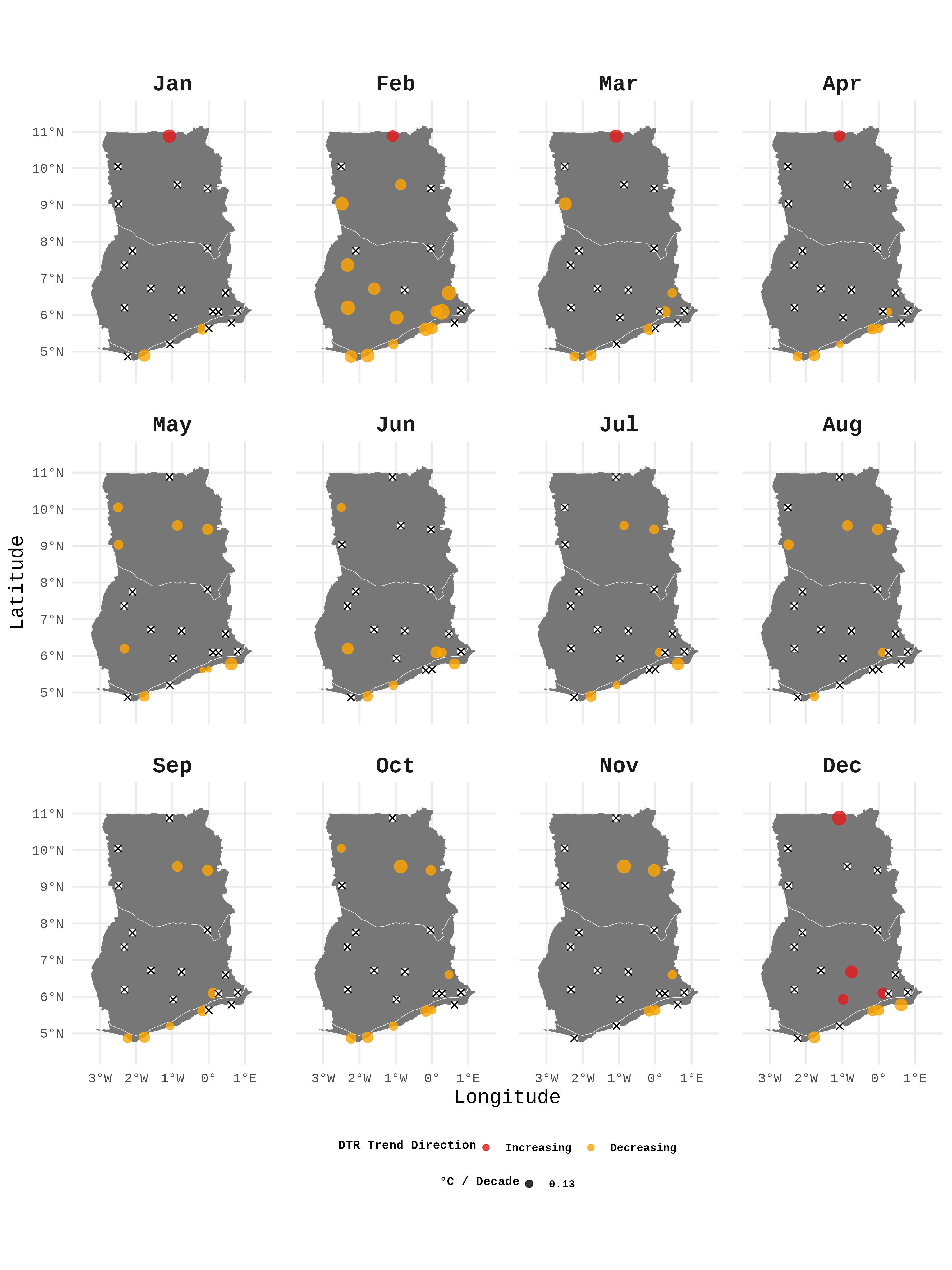}
	\caption{Mean monthly diurnal temperature range (DTR) trends in Ghana. Increasing trends are shown with red circles, while decreasing trends are in orange. Black X on fixed white circles indicate the trend at the location is not significant (p $\geq$ 0.05). Circle size is proportional to the magnitude of change}
	\label{fig:monthly_dtr_trends}
\end{figure}

Figure~\ref{fig:monthly_dtr_trends} presents the monthly DTR trends for the period 1983--2021 at the various stations. The results showed that the trends in DTR were more nuanced in the different months. February revealed significant downward trends at most of the stations in the Forest and Coastal zones, while May revealed significant downward trends at most stations in the Savannah zone. Navrongo was seen to have significant upward trends in December and January to April, and non-significant trends in the other months. Akuse, Abetifi, and Akim Oda showed significant increasing trends in December, while sometimes with significant decreasing trends or had no significant trends in other months. These results further highlight the highly localised nature of temperature trends, as well as the asymmetric nature of the Tmax and Tmin warming in Ghana.      

\section{Discussion}\label{sec:discussion}
The primary objective of this research was to determine whether temperature trends in Ghana are localised or if they align with broader regional patterns established by traditional zonal averaging approaches. A secondary objective involved analysing the monthly evolution of these trends. By analysing a network of 22 meteorological stations, this study provides several insights into the evolving thermal landscape of the country.

One key finding is that temperature trends in Ghana are highly localised, which is consistent with the spatially diverse patterns identified in coastal regions by \cite{Ankrah2023}. The specific manifestations of these trends varied considerably in both space and time (Figures~\ref{fig:monthly_tmax_trends}--\ref{fig:monthly_trends_wa} and \ref{fig:monthly_dtr_trends}). In general, annual Tmax and Tmin exhibited significant increasing trends across most locations, aligning with broader national assessments \citep{Atiah2021, Oduro2021}, as well as site-specific studies in the coastal zone \citep{Ankrah2023}, the north-east \citep{Asamoah2020}, and the Asante Akim North District \citep{BaffourAta2023}. While \cite{Abbam2018} reported non-significant Tmin trends in the Upper East Region, which is consistent with the findings for Navrongo in our study, some nuanced differences emerged regarding the Upper West Region. Specifically, while \cite{Abbam2018} found no significant Tmin trends for the region as a whole, our study identified significant increasing trends at the Wa and Bole stations, located in this region. Such discrepancies may be attributed to the use of gridded data at a resolution of 0.5$^\circ$ by 0.5$^\circ$ in their work, which potentially smoothed the finer details captured by the point observations used in our work.

The seasonal analysis indicated that significance and magnitude of trends vary throughout the year, further refining the findings of \cite{Oduro2021} regarding upward trends during the February to September periods. Our study found that Tmax increased most rapidly during December and January. The seasonality of Tmin was more complex, though December, January, and February emerged as the fastest-warming months for the majority of stations, particularly in the Forest and Coastal zones. These observations are in close agreement with \cite{Iddrisu2025}, who identified variable monthly trends, especially for Tmax, in the Tamale Metropolis. Minor differences in the significance of specific monthly trends, such as those observed for March and June in Tamale, may arise from the fact that \cite{Iddrisu2025} used an average of five stations, whereas our study focused on individual station dynamics.

Beyond spatial and temporal variations, this research highlights a distinct diurnal asymmetry, with minimum temperatures generally warming at a faster rate than maximum temperatures. Significant downward trends in the mean annual DTR were observed at nearly all stations within the coastal zone (Figure~\ref{fig:annual_dtr_trends}), which corroborates the results of \cite{Ankrah2023}. DTR is known to be a direct indicator of climate change \citep{Cox2020, Ankrah2023, Mall2021}. The coastal cities of Ghana are marked by urbanisation and increased industrial activities. This may be a contributing factor to the trend, as heat through urbanisation and population growth has been found to decrease DTR \citep{Kalnay2003, Mohan2015}. The monthly analysis further revealed localised DTR shifts in the Savannah and Forest zones, where significant upward and downward trends were detected (Figure~\ref{fig:monthly_dtr_trends}). This suggests the influence of site-specific factors, including cloud cover \citep{Cox2020}, localised emissions, and changes in land-use \citep{Adeyeri2026, Hardwick2015, li2018leaf, Easterling1997}. Such diurnal asymmetry may have serious socio-economic implications, as accelerated nocturnal warming can disrupt crop maturation and increase thermal stress in both human populations and broader biological systems \citep{Cox2020}.

\section{Conclusion}\label{sec:conclusion}
This study was carried out with two primary objectives: 1. to evaluate whether temperature trends in Ghana are localised or if they align with broader regional patterns established by traditional zonal averaging; and 2. to analyse the monthly evolution of these trends. Using daily Tmax and Tmin observations, subjected to rigorous quality control, homogeneity testing, and homogenisation against AgERA5 reanalysis data, this research provides a robust empirical foundation to understand the changing thermal landscape of the country.

The results demonstrate that temperature trends in Ghana are highly localised, with significant spatial heterogeneity that is often masked by broader regional averages. Furthermore, analysis of the monthly evolution of these trends revealed a distinct seasonality in the warming rates, with Tmax increasing most rapidly during December and January, while the warming of Tmin was most pronounced during the months of December through February. These findings underscore a critical need for high-resolution spatio-temporal monitoring. By contextualising climate impacts at a granular level, policymakers can move beyond generalised responses toward bespoke, site-specific solutions tailored to the unique vulnerabilities and seasonal risks of each locality.

The clear evidence of increasing temperatures across all studied locations, combined with the observed asymmetric warming in diurnal trends, particularly the narrowing of the DTR in coastal regions, makes the implementation of integrated climate policies imperative. Priority should be given to large-scale afforestation programmes to restore natural carbon sinks and alleviate localised heat effects, alongside an accelerated transition towards renewable energy to ensure long-term sustainability. By embedding these site-specific and seasonal climatic insights into national policy frameworks, Ghana can better equip its communities to navigate the complexities of a warming climate while fostering a resilient, climate-smart economy.

Future research can build upon this framework by applying the proposed high-resolution methodology to a broader range of climatic variables and an expanded network of meteorological stations nationwide. Such an expansion would facilitate a more comprehensive understanding of multi-dimensional climate shifts and further refine the precision of regional adaptation strategies.

\backmatter

\bmhead{Acknowledgments}

The authors acknowledge and thank the Ghana Meteorological Agency for providing station data for the study to be conducted. We also thank all reviewers for their comments to improve this manuscript.

\section*{Declarations}

\begin{itemize}
	\item \textbf{Funding} This publication was made possible by a grant from Carnegie Corporation of New York (provided through the African Institute for Mathematical Sciences). The statements made and the views expressed are solely the responsibility of the authors.
	
	% \item \textbf{Competing interests} The authors have no relevant financial or non-financial interests to disclose.
	
	% \item \textbf{Ethics approval} Not applicable 
	% \item \textbf {Consent to participate} Not applicable
	
	% \item \textbf{Consent for publication} Not applicable
	
	% \item \textbf{Data Availability} The AgERA5 data is publicly available. The station data can be obtained from the Ghana Meteorological Agency. 
	
	% \item \textbf{Code availability} Code is available upon reasonable request.
	
	% \item \textbf{Authors' contributions} 
	% Conceptualisation: J.B., D.S.; Methodology: J.B., D.S.; Formal analysis: J.B.; Data curation: J.B., D.S.; Writing --- original draft: J.B.; Writing --- review \& editing: J.B., D.S., D.N.; Visualisation: J.B.; Supervision: D.S., D.N.
	
\end{itemize}

\noindent
%If any of the sections are not relevant to your manuscript, please include the heading and write `Not applicable' for that section. 

\bigskip
\iffalse
\begin{flushleft}%
Editorial Policies for:

\bigskip\noindent
Springer journals and proceedings: \url{https://www.springer.com/gp/editorial-policies}

\bigskip\noindent
Nature Portfolio journals: \url{https://www.nature.com/nature-research/editorial-policies}

\bigskip\noindent
\textit{Scientific Reports}: \url{https://www.nature.com/srep/journal-policies/editorial-policies}

\bigskip\noindent
BMC journals: \url{https://www.biomedcentral.com/getpublished/editorial-policies}
\end{flushleft}
\fi

\begin{appendices}
	
	\section{Homogeneity tests definitions}\label{secA1}

	\subsection{The Standard Normal Homogeneity Test (SNHT)}
	
	The Standard Normal Homogeneity Test, as developed by \citet{alexandersson1986homogeneity}, utilizes a likelihood ratio statistic to compare the mean of the first $m$ years with the mean of the remaining $n-m$ years in a given series $Y$. The test statistic $T(m)$ is defined as follows:
	
	\begin{equation}
		T(m) = m\bar{z}_1^2 + (n - m)\bar{z}_2^2, \quad m = 1,\dots,n
	\end{equation}
	
	In this equation, the standardized mean values for the two sub-periods are calculated as:
	
	\begin{equation}
		\bar{z}_1 = \frac{1}{m}\sum_{i=1}^{m}(Y_i - \bar{Y})/\sigma \quad\text{and}\quad \bar{z}_2 = \frac{1}{n-m}\sum_{i=m+1}^{n}(Y_i - \bar{Y})/\sigma
	\end{equation}
	
	where $n$ is the number of observations in the series, and $\bar{Y}$ and $\sigma$ represent the mean and sample standard deviation of the entire series, respectively. When a break occurs at year $M$, the function $T(m)$ reaches its maximum value near $m = M$. The global test statistic $T$ is determined by finding the maximum value across the sequence:
	
	\begin{equation}
		T = \max_{1\leq m < n} T(m)
	\end{equation}
	
	The SNHT is a parametric test that assumes the underlying data follows a normal distribution. A primary property of this test is its high sensitivity to shifts occurring at the beginning or the end of a time series. While it is highly effective at detecting abrupt changes in mean, its parametric nature makes it less robust to outliers or non-normal data distributions compared to rank-based methods.
	
	\subsection{Pettitt Test (PT)}
	The Pettitt test \citep{pettitt1979non} is a non-parametric approach for change point detection based on the ranks of the observations. This method is particularly useful when the distribution of the data is unknown or non-normal. For a series $Y_1, \dots, Y_n$, let $r_1, \dots, r_n$ be the corresponding ranks. The test statistics are computed as follows:
	
	\begin{equation}
		U_m = 2\sum_{i=1}^{m} r_i - m(n + 1), \quad m = 1,\dots,n
	\end{equation}
	
	If a change point occurs at year $M$, the statistic $U_m$ reaches either its maximum or minimum value near $m = M$. The final test statistic is the maximum absolute value of $U_m$:
	
	\begin{equation}
		U_M = \max_{1\leq m \leq n} |U_m|
	\end{equation}
	
	The Pettitt test is recognized for its robustness against outliers because it utilizes ranks rather than the absolute values of the data. Unlike the SNHT, this test is more sensitive to shifts occurring in the middle of a time series and is generally less powerful at identifying changes located near the boundaries of the data record.
	
	\subsection{Buishand's tests}
	The assessment of homogeneity in the climate series is based on the calculation of adjusted partial sums and specific test statistics designed to identify shifts in the mean. The adjusted partial sums, denoted as $S_m^*$, represent the cumulative deviations of the observations from the series mean \citep{buishand1982some}. These are defined as:
	
	\begin{equation}
		S_0^* = 0, \quad S_m^* = \sum_{i=1}^{m}(Y_i - \bar{Y}), \quad m = 1, \dots, n
	\end{equation}
	
	In this expression, $Y_i$ represents the observation at time $i$ and $\bar{Y}$ denotes the sample mean. In a homogeneous series, these sums fluctuate around zero, while a sudden shift in the mean causes the sums to trend upward or downward, typically reaching a peak or trough at the point of change. 
	
	The internal variability of the dataset is accounted for by the sample standard deviation, $\sigma$ %, which is calculated as follows:
	
	% \begin{equation}
	% D = \sqrt{\frac{1}{n} \sum_{i=1}^{n}(Y_i - \bar{Y})^2}
	% \end{equation}
	
	\subsubsection{Buishand likelihood ratio test (BLRT)}
	To detect an abrupt change at an unknown time, the likelihood ratio statistic $V$ is employed \citep{Buishand1984}. This statistic is defined as:
	
	\begin{equation}
		V = \max_{1 \le m \le n-1} \left\{ \frac{|S_m^*|}{\sigma \{m(n-m)\}^{1/2}} \right\}
	\end{equation}
	
	The $V$ statistic is particularly effective for identifying shifts that occur near the beginning or the end of a time series \citep{Buishand1984}. This sensitivity is due to the weighting factor in the denominator, which emphasizes deviations located toward the boundaries of the data series \citep{Buishand1984}.
	
	\subsubsection{Buishand's $U$ test (BUT)}
	Alternatively, the Bayesian statistic $U$ is used to evaluate shifts when the position of the change point is assumed to have a uniform prior distribution \citep{Buishand1984}. The statistic is defined as:
	
	\begin{equation}
		U = \frac{1}{n(n+1)} \sum_{m=1}^{n-1} \left( \frac{S_m^*}{\sigma} \right)^2
	\end{equation}
	
	Unlike the likelihood ratio test, the $U$ statistic gives less weight to the series endpoints and is therefore more sensitive to mean shifts occurring in the middle of the sequence \citep{Buishand1984}.

%	\section{Homogeneity test results on annual DTR series}\label{secA2}
	
	\begin{figure}[H] 
		\centering    \includegraphics[width=0.95\textwidth]{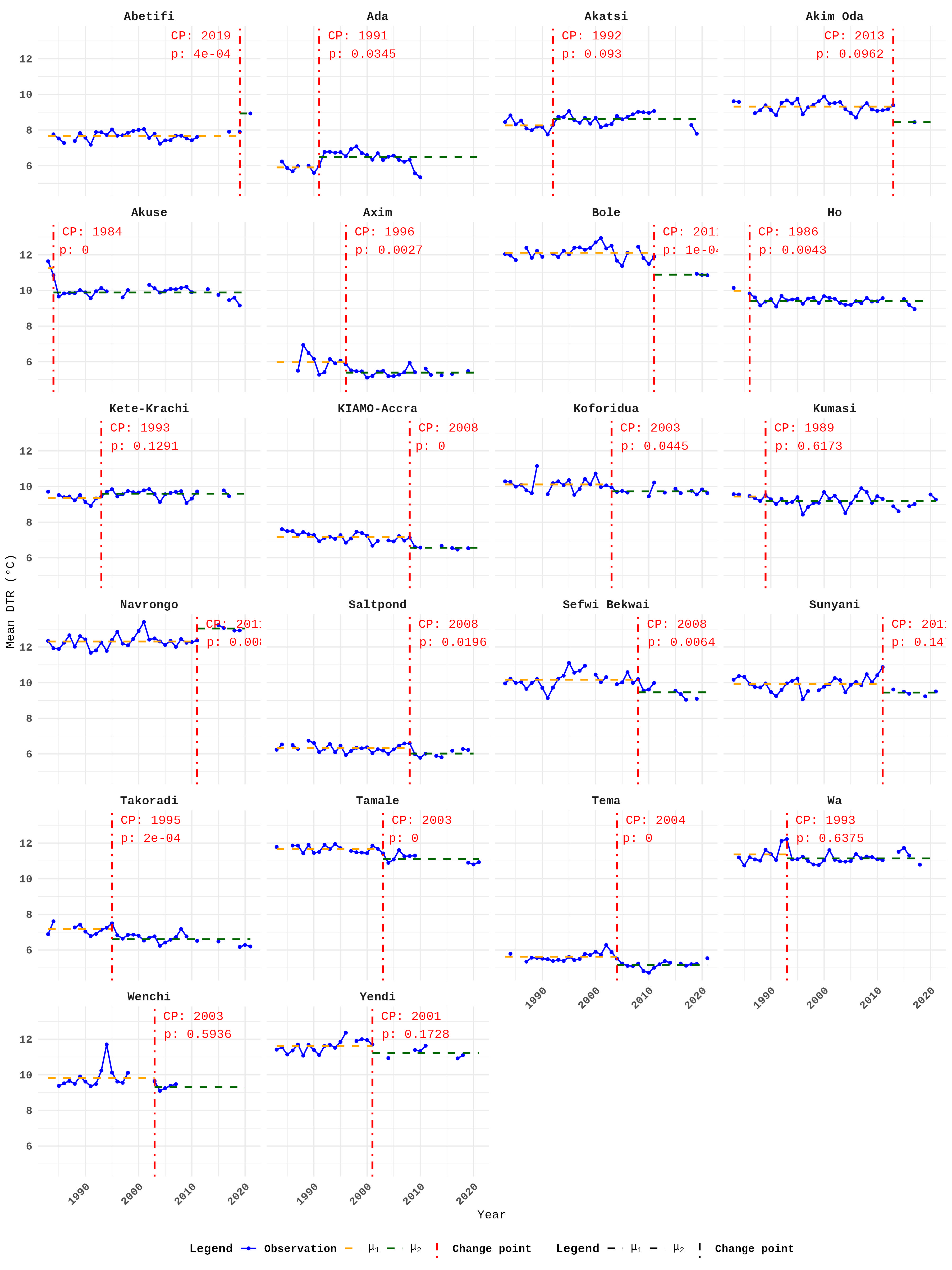}
		\caption{SNHT results}
		\label{fig:snht_results}
	\end{figure}
	
	\begin{figure}[H] 
		\centering    \includegraphics[width=0.95\textwidth]{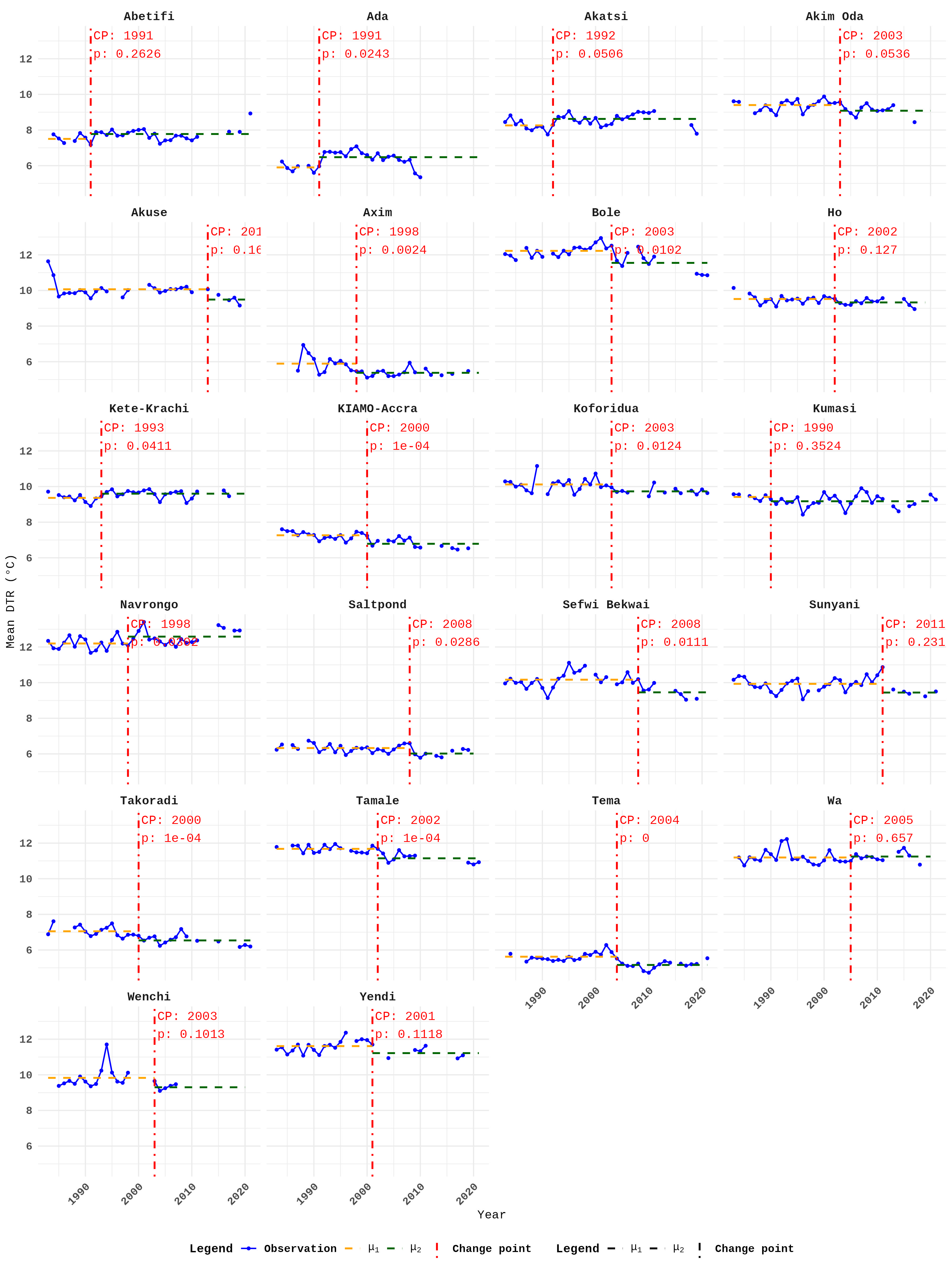}
		\caption{Pettitt test results}
		\label{fig:pt_results}
	\end{figure}
	
	\begin{figure}[H] 
		\centering    \includegraphics[width=0.95\textwidth]{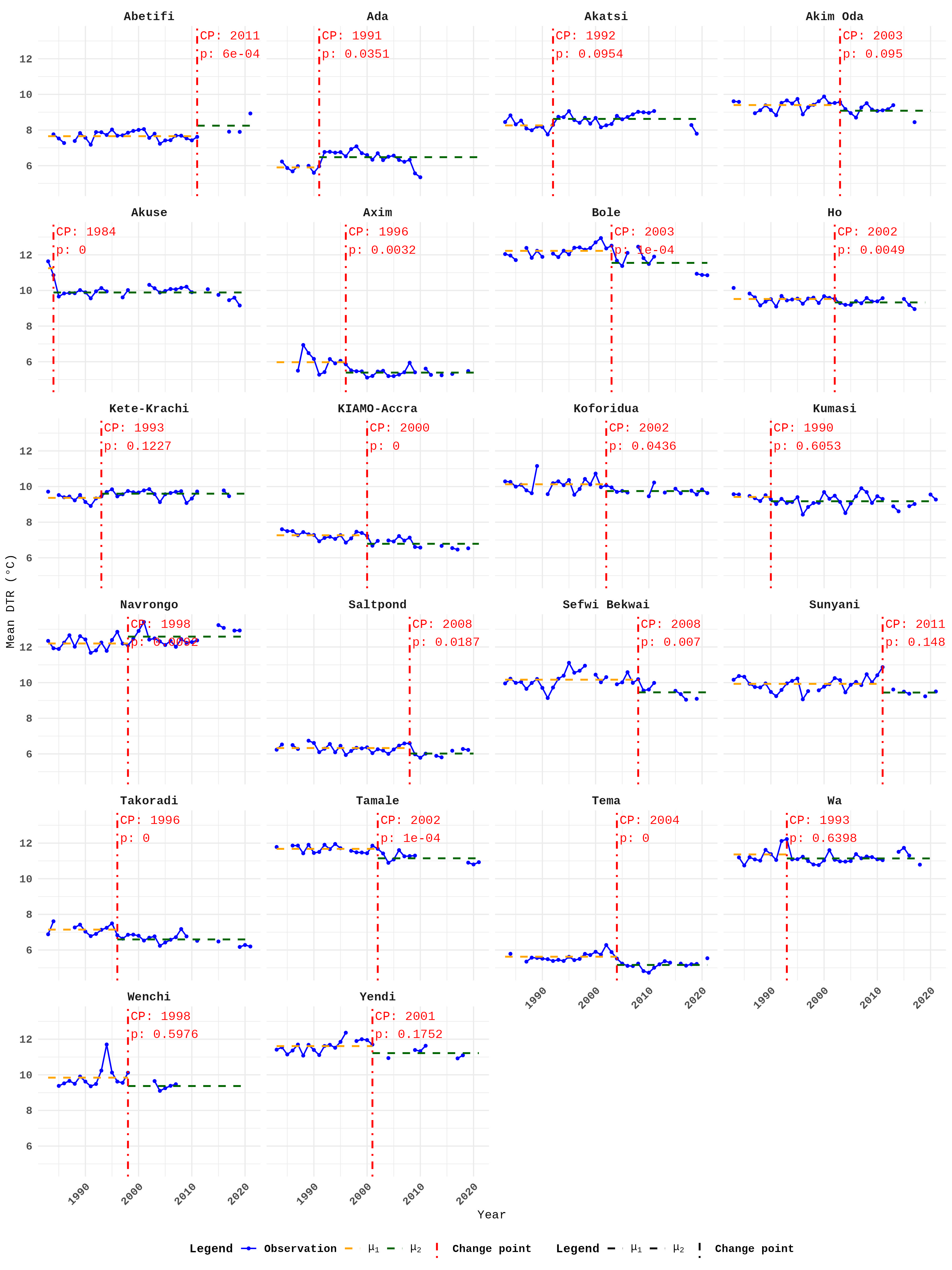}
		\caption{BLRT results}
		\label{fig:blrt_results}
	\end{figure}
	
	\begin{figure}[H] 
		\centering    \includegraphics[width=0.95\textwidth]{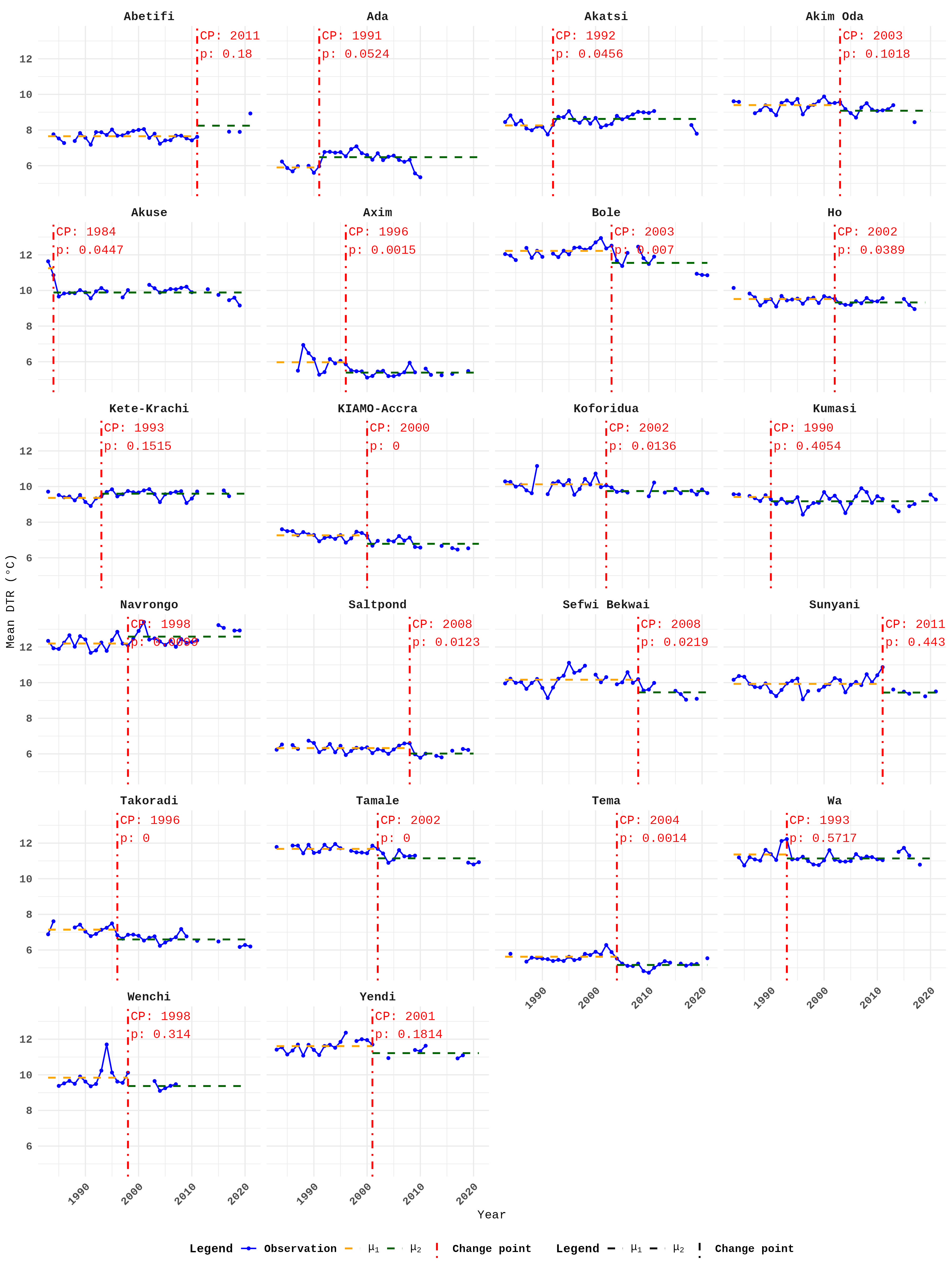}
		\caption{BUT results}
		\label{fig:but_results}
	\end{figure}
	
\end{appendices}


\bibliography{sn-bibliography}% common bib file
%\input sn-article.bbl

\end{document}